\documentclass[prb,twocolumn,showpacs,preprintnumbers,amsmath,amssymb]{revtex4}
\usepackage{graphicx}% Include figure files
\usepackage{dcolumn}% Align table columns on decimal point
\usepackage{bm}% bold math

\begin{document}

\preprint{APS/123-QED}

\title{Anisotropic magnetoresistance in a 2DEG in a quasi-random magnetic field}

\author{A. W. Rushforth}
 \altaffiliation[Present address ]{Cavendish Laboratory, Madingley Road, Cambridge CB3 0HE, England.}%Lines break automatically or can be forced with \\
 \email{awr1001@cam.ac.uk}
\author{B. L. Gallagher}
\author{P. C. Main}
\author{A. C. Neumann}
\author{M. Henini}
 \affiliation{School of Physics and Astronomy, The University of Nottingham, University Park, Nottingham NG7 2RD, England
}

\author{C. H. Marrows}
\author{B. J. Hickey}
\affiliation{Department of Physics, The University of Leeds, Leeds LS2 8JT, England.
}%

\date{\today}% It is always \today, today,

\begin{abstract}
We present magnetotransport results for a 2D electron gas (2DEG) subject to the quasi-random magnetic field produced by randomly positioned sub-micron Co dots deposited onto the surface of a GaAs/Al$_{x}$Ga$_{1-x}$As heterostructure. We observe strong local and non-local anisotropic magnetoresistance for external magnetic fields in the plane of the 2DEG. Monte-Carlo calculations confirm that this is due to the changing topology of the quasi-random magnetic field in which electrons are guided predominantly along contours of zero magnetic field.
\end{abstract}

\maketitle

The transport properties of a two-dimensional electron gas (2DEG), subject to a spatially random magnetic field, have attracted great theoretical interest recently \cite{RefA1,RefA2}. This is partly due to its relevance to the fractional quantum Hall effect, which can be understood in terms of composite fermions moving in an effective random magnetic field \cite{RefA2}. The problem of a 2DEG subject to large amplitude random magnetic fields, with correlation lengths that are small compared to the electron mean free path, is particularly interesting because the transport properties are predicted to be dependent on semi-classical "snake orbits" trajectories in which electrons are guided along lines of zero magnetic field \cite{RefA2}. In a previous study we demonstrated that such "snake orbits" can lead to very large magnetoresistances when a 2DEG is subject to large amplitude periodic magnetic fields \cite{RefA3}. In this paper we show that electrons are indeed guided along contours of zero magnetic field in a quasi-random magnetic field and that this gives rise to a new type of anisotropic magnetoresistance. 

The devices investigated in this study are hybrid ferromagnetic/ semiconductor structures in which the magnetic field at the 2DEG is produced by randomly positioned sub-micron Co dots fabricated on the surface of a GaAs based heterostructure. Recently, there has been considerable interest in the properties of hybrid ferromagnetic/ semiconductor devices due to their potential applications as magnetic field sensors \cite{RefA4} and magnetic storage devices \cite{RefA5}. There have been several previous experimental studies of a 2DEG subject to a random magnetic field. However, these concerned very weak random magnetic fields \cite{RefE1}, or random magnetic fields with correlation lengths approximately equal to \cite{RefE2} or much larger than \cite{RefE3} the electron mean free path. Recently, we showed how large amplitude, short correlation length random magnetic fields can be realised by using the naturally occurring domain structures of CoPd films \cite{RefE4}. The method considered here also produces a large amplitude, short correlation length, random magnetic field, but has the particular advantage that the topology of the field can be controlled by an external in-plane magnetic field. 

Our device consists of a GaAs/Al$_{x}$Ga$_{1-x}$As heterostructure containing a 2DEG 35nm below the surface. At 1.3K the electron density, $n=3.8\times 10^{15}m^{-2}$ and the mobility, $\mu=45m^{2}V^{-1}s^{-1}$ corresponding to a mean free path, $\lambda=4.5\mu m$ \cite{RefF}. The device has a standard Hall bar design as shown in Fig. 1(a). The random magnetic field is produced by depositing a randomly distributed pattern of Co dots onto the surface of the Hall bar. This is represented by the shaded area in the figure. The Co dots have a diameter of 500nm and a thickness of 70nm. The pattern was produced using electron beam lithography and dc magnetron sputtering to deposit the Co. A computer program was used to generate the positions of the dots which cover 50\%\ of the surface area and are distributed so that they can touch each other, but not overlap. Figure 1(b) shows a scanning electron microscope image of the Co dots. There are some areas where the Co has remained between a cluster of dots after lift-off, but such occurrences are rare.

\begin{figure}
\includegraphics{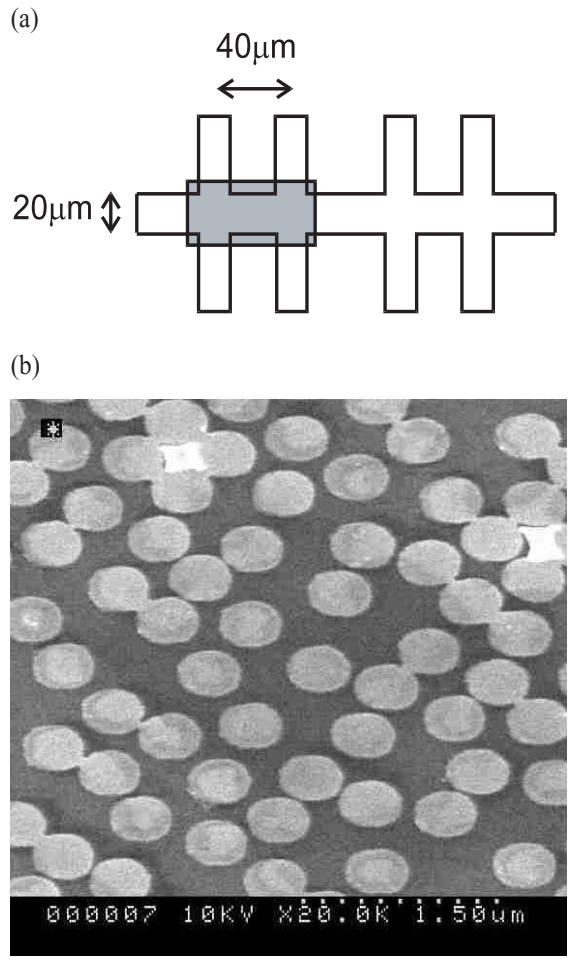}% Here is how to import EPS art
\caption{\label{fig1:epsart}(a) Standard Hall bar design of the device. The shaded area represents the region where the Co dots are deposited. (b) A scanning electron microscope image of the Co dots pattern.}
\end{figure}

\begin{figure}
\includegraphics{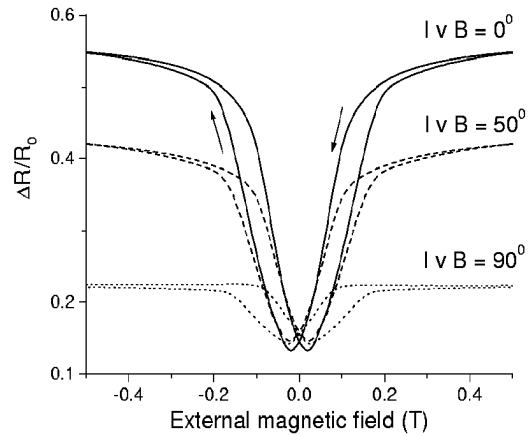}% Here is how to import EPS art
\caption{\label{fig2:epsart} The longitudinal magnetoresistance measured at 1.3K with the magnetic field applied parallel to the plane of the 2DEG. Results are shown for angles of $0^{\circ}$, $50^{\circ}$ and $90^{\circ}$ between the current and the applied field. The arrows show the direction that the magnetic field is sweeping. }
\end{figure}

Figure 2 shows the longitudinal magnetoresistance of the covered section of the device measured at 1.3K with a magnetic field applied parallel to the plane of the 2DEG \cite{RefG}. The magnetoresistance is defined as $\Delta R/R_0$=$(R-R_0)/R_0$, where $R$ is the measured resistance of the covered section when the magnetic field is applied and $R_0$ is the resistance of the uncovered section in zero field. Results are shown for angles of $0^{\circ}$, $50^{\circ}$ and $90^{\circ}$ between the current along the Hall bar and the direction of the applied field. We observe a positive magnetoresistance for all angles which saturates at approximately $\pm 0.2$~T. The size of the magnetoresistance is largest when the field is applied in the direction of the current and decreases as the angle between the field and the current is rotated to $90^{\circ}$. The presence of hysteresis in the measurements is consistent with the fact that sputtered Co films tend to have the easy axis of magnetisation in the plane \cite{RefH}. Indeed, we carried out transport measurements with the magnetic field applied perpendicular to the device and no hysteresis was observed.

\begin{figure}
\includegraphics{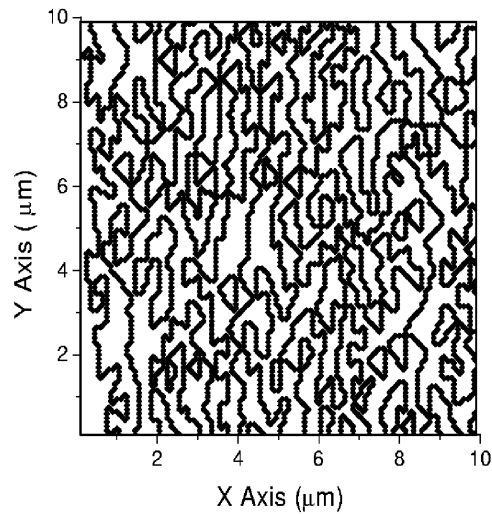}% Here is how to import EPS art
\caption{\label{fig3:epsart} The contours of zero magnetic field ($B_{z}=0T$) taken from the calculated magnetic field profile at the site of the 2DEG due to a random array of Co dots. The magnetisation of the Co dots is along the x-axis. }
\end{figure}

We have shown previously that the resistance of a 2DEG increases with increasing random magnetic field amplitude \cite{RefE4}. The magnetisation state of the dots is not known at zero external magnetic field. However, we expect the formation of multidomain or vortex states, as has been observed by Miramond et al \cite{RefI} for dots of diameter greater than 50nm. Such states have a high degree of, but not complete, flux closure and the small net stray field will produce a small amplitude random magnetic field at B=0T. This is the origin of the positive value of $\Delta R/R_0$ at B=0T. With increasing external field, domains aligned with the applied field will grow until eventually, the magnetisation of the dots is saturated in the direction of the applied field. The amplitude of the random magnetic field at the 2DEG will therefore grow with increasing external field until saturation magnetisation is reached. This will lead to the observed increase and then saturation of the measured resistance. 

We now restrict ourselves to analysing the resistance when the magnetisation of the dots is saturated. The magnetic field profile at the site of the 2DEG due to the random array of Co dots can be calculated numerically, since we know the positions of all the Co dots. The dynamics of the 2D electrons are only sensitive to $B_{z}$, the component of the magnetic field perpendicular to the 2DEG. We have calculated $B_{z}$ for our device \cite{RefJ} and find that the rms amplitude is 0.13T at saturation. Figure 3 shows the contours of zero field ($B_{z}=0T$) calculated for the case when the magnetisation of the dots is saturated in the x-direction. It can be seen that the contours are preferentially aligned in the direction perpendicular to the magnetisation of the dots. Using the calculated magnetic field profile we have calculated the expected device resistance using the semi-classical Kubo formalism. The method involves calculating the trajectories of electrons with the Fermi velocity in the random magnetic field profile. By averaging the velocity-velocity correlation function over all of the trajectories the diffusivity tensors for the system can be calculated using the Kubo formula \cite{RefK}:
\begin{equation}
D_{xx} = \int_{0}^{\infty} \langle v_{x}(0)v_{x}(t) \rangle dt
\label{eq:one}
\end{equation}

Here the x-axis is parallel to the direction of the applied field. We can then calculate the magnetoresistance \cite{RefJ}. Scattering is included in the model by randomising the electron direction using a Monte-Carlo method with appropriate probability distributions for the scattering angle and scattering time. Figure 4 shows the calculated final positions of 10,000 electrons starting at the same initial position (0,0) with the Fermi velocity and initial angles spread evenly over $360^{\circ}$. Each electron is allowed to travel 6 momentum relaxation mean free path lengths. The results show that the electrons are preferentially guided in the direction of the $B_{z}=0 T$ contours. Our calculations show that this leads to an enhancement of the ratio $D_{yy}/D_{xx}$ as is also the case in a sign alternating magnetic field profile with contours of zero $B_{z}$ aligned in the y-direction \cite{RefA3}.

\begin{figure}
\includegraphics{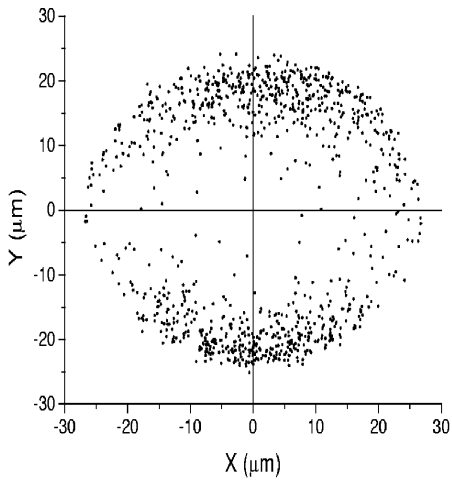}% Here is how to import EPS art
\caption{\label{fig4:epsart} The final positions of 10,000 electrons starting at the coordinates (0,0) with the Fermi velocity and initial angles spread evenly over $360^{\circ}$. The electron trajectories were calculated within the field profile used to produce Fig. 3. }
\end{figure}

\begin{figure}
\includegraphics{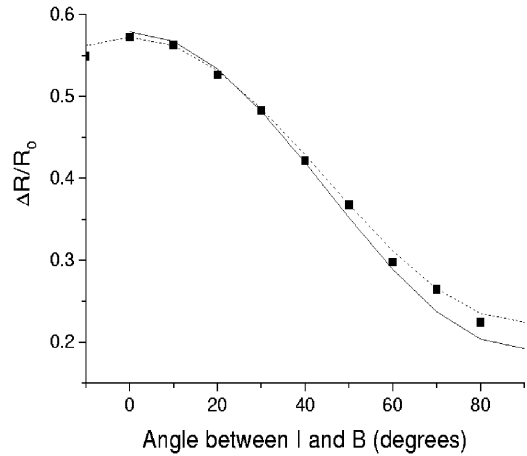}% Here is how to import EPS art
\caption{\label{fig5:epsart} The magnetoresistance, measured with an in-plane field of 0.5T, plotted against the angle between the current and the magnetic field. The dashed line is a fit of equation 2 to the data using the measured values of $\rho_{\perp}$ and $\rho_{\parallel}$. The solid line shows the results of the numerical calculations. }
\end{figure}

Figure 5 shows the magnetoresistance, measured for an in-plane field of 0.5T, against the angle between the current and the applied field. If the anisotropic magnetoresistance is due to only the induced magnetisation and the samples have no additional intrinsic anisotropy, arising from anisotropy in the dot positions for example, then one expects \cite{RefM}  
\begin{equation} \rho_{xx}=\rho_{\perp}+(\rho_{\parallel}-\rho_{\perp})cos^{2}{\theta}
\label{eqn:two}
\end{equation}
 where $\rho_{\parallel}$ is the measured resistivity for $\theta$=$0^{\circ}$ and $\rho_{\perp}$ is the measured resistivity for $\theta$=$90^{\circ}$. The dashed line in Fig. 5 shows that this gives an excellent fit to our data. Also shown are the results of numerical calculations of the conductivity tensors. The fit was obtained by using the thickness of the Co dots as the only adjustable parameter. We see excellent quantitative agreement to our data when a thickness of 75nm is used, which is in very close agreement to the nominal thickness of 70nm. 

Another clear example of the effect of contours of $B_{z}=0T$ guiding the electron motion can be observed by performing measurements in a non-local geometry. The inset to Fig. 6 shows the experimental arrangement. A constant ac current of $300\mu A$ is passed between two voltage probes in the y-direction, across the Hall bar, and the voltage is measured between two adjacent voltage probes. The external magnetic field is applied parallel to the plane of the 2DEG. Figure 6 shows the measured voltage, as the external magnetic field is swept to $\pm 0.5T$, for angles of $0^{\circ}$, $45^{\circ}$ and $90^{\circ}$ between the current and the field. When the magnetic field is parallel to the current, the contours of $B_{z}=0T$ will guide electrons in the x-direction, perpendicular to the direction of net current flow i.e. the current will tend to spread along the Hall bar. This results in an increase in the measured voltage. When the magnetic field is perpendicular to the current, the contours of $B_{z}=0T$ will tend to focus electrons along the y-direction reducing the spread of the current density in the x-direction. This results in a decrease in the measured voltage. For $45^{\circ}$ one expects the existence of the orientated zero field contours to have no effect. Therefore, the form of the measured voltages in Fig. 6 is consistent with the picture of electron trajectories being guided along zero field contours.

In conclusion, we have observed a new type of anisotropic magnetoresistance and demonstrated that this arises from the dependence of the anisotropy of the quasi-random magnetic field on the direction of the net magnetisation. In particular, we have shown the strong influence of the contours of zero magnetic field, which has been predicted theoretically \cite{RefA2}. 

\begin{acknowledgments}
We gratefully acknowledge funding by the EPSRC (UK) and the ESPRIT SPIDER programme (EU). We also wish to thank S.Thoms for allowing us to use the electron beam facility at the Department of Electronics and Electrical Engineering, The University of Glasgow.
\end{acknowledgments}

\begin{figure}
\includegraphics{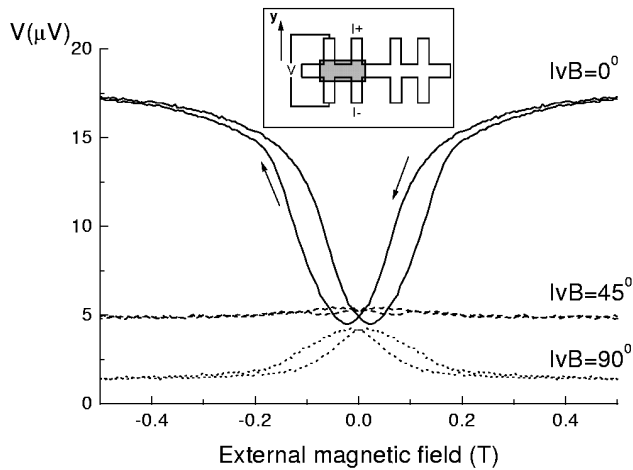}% Here is how to import EPS art
\caption{\label{fig6:epsart} The inset shows the experimental arrangement for the non-local measurement. The measured voltage is shown for the external magnetic field applied in the plane of the 2DEG at angles of $0^{\circ}$ (solid curve), $45^{\circ}$ (dashed curve)  and $90^{\circ}$ (dotted curve) to the direction of the current. }
\end{figure}


\bibliography{apssamp}% Produces the bibliography via BibTeX.

\begin{thebibliography}{}



\bibitem{RefA1}
J.T. Chalker, D.G. Polyakov, F. Evers, A.D. Mirlin, and P. W\" olfle, Phys. Rev. B \textbf{66}, 161317 (2002).
\bibitem{RefA2}
F. Evers, A.D. Mirlin, D.G. Polyakov, P. W\" olfle, Phys. Rev. B \textbf{58}, 15321 (1998);
J. Wilke, A.D. Mirlin, D.G. Polyakov, F. Evers and P. W\" olfle, Phys. Rev. B \textbf{61}, 13774 (2000) and references therein.
\bibitem{RefA3}
A. Nogaret, S. Carlton, B.L. Gallagher, P.C. Main, M. Henini, R. Wirtz, R. Newbury, M.A. Howson, S.P. Beaumont, Phys. Rev. B \textbf{55}, 16037 (1997).
\bibitem{RefA4}
B.L. Gallagher, V. Kubrak, A.W. Rushforth, A.C. Neumann, K.W. Edmonds, P.C. Main, M. Henini, C.H. Marrows, B.J. Hickey and S. Thoms, Physica E \textbf{11}, 171 (2001); V. Kubrak, A. Neumann, B.L. Gallagher, P.C. Main, M. Henini, C.H. Marrows, B.J. Hickey, J. Appl. Phys. \textbf{87}, 5986 (2000); A.K. Geim, S.V. Dubonos, J.G.S. Lok, I.V. Grigorieva, J.C. Maan, L. Theil Hansen and P.E. Lindelof, Appl. Phys. Lett. \textbf{71}, 2379 (1997).
\bibitem{RefA5}
M. Johnson, B.R. Bennett, M.J. Yang, M.M. Miller and B.V. Shanabrook, Appl. Phys. Lett. \textbf{71}, 974 (1997).
\bibitem{RefE1}
A.K. Geim, S.J. Bending, I.V. Grigorieva and M.G. Blamire, Phys. Rev. B \textbf{49}, 5749 (1994); A.K. Geim, S.J. Bending and I.V. Grigorieva, Phys. Rev. Lett. \textbf{69}, 2252 (1992); P.D. Ye, D. Weiss, G. Lutjering, R.R. Gerhardts, K. v. Klitzing and K. Eberl, Proceedings of the 23rd International Conference on the Physics of Semiconductors., pp 1537; A. Smith, R. Taboryski, L.T. Hansen, C.B. S\o rensen, P. Hedegard and P.E. Lindelof, Phys. Rev. B \textbf{50}, 14726 (1994); L.T. Hansen, R. Taboryski, A. Smith, P.E. Lindelof and P. Hedegard, Surf. Sci. \textbf{362}, 349 (1996). 
\bibitem{RefE2}
G.M. Gusev, A.A. Quivy, J.R. Leite, A.A. Bykov, N.T. Moshegov, V.M. Kudryashev, A.I. Toropov and Y.V. Nastaushev, Semicond. Sci. Tech. \textbf{14}, 1114 (1999).
\bibitem{RefE3}
F.B. Mancoff, R.M. Clarke, C.M. Marcus, S.C. Zhang, K. Campman and A.C. Gossard, Phys. Rev. B \textbf{51}, 13269 (1995).
\bibitem{RefE4}
A.W. Rushforth, B.L. Gallagher, P.C. Main, A.C. Neumann, C.H. Marrows, I. Zoller, M.A. Howson, B.J. Hickey and M. Henini, Physica E \textbf{6}, 751 (2000).
\bibitem{RefF}
Obtained from measurements of the Hall and Shubnikov de-Haas effects.
\bibitem{RefG}
Note that we expect the transport properties of the 2DEG to be dependent on only the perpendicular component of the magnetic field, therefore the external field should not affect the 2DEG directly. 
\bibitem{RefH}
O. Kohmoto and T. Yamamoto, J. Mag. Mag. Mat. \textbf{71}, 33 (1987); H. Qui, T. Ohbuchi, H. Nakai and M. Hashimoto, Appl. Surf. Sci. \textbf{92}, 47 (1996); O. Kohmoto, Phys. Stat. Sol. \textbf{176}, 1039 (1999).
\bibitem{RefI}
C. Miramond, C. Fermon, F. Rousseaux, D. Decanini and F. Carcenac, J. Mag. Mag. Mat. \textbf{165}, 500 (1997).
\bibitem{RefJ}
A.W. Rushforth, Ph.D. Thesis (2000).
\bibitem{RefK}
C.W.J. Beenakker and H. Vanhouten, "Quantum Transport in Semiconductor Nanostructures," Solid State Physics Advances in Research and Applications, \textbf{44}, 1-228, (1991).
\bibitem{RefM}
T.R. McGuire and R.I. Potter, IEEE Trans. Magn. MAG-11, 618 (1973).

\end{thebibliography}

\end{document}